# Infinite-dimensional Bäcklund transformations between isotropic and anisotropic plasma equilibria.

# Infinite symmetries of anisotropic plasma equilibria.

**Alexei F. Cheviakov**

**Queen's University at Kingston, 2002.**

Research advisor: Professor O. I. Bogoyavlenskij.

*In this paper we establish Bäcklund transformations between solutions of several cases of classical isotropic MHD and plasma equilibria and corresponding anisotropic equilibria.*

*The transformations appear to be infinite-dimensional and allow building solutions with variety of physical properties.*

*We also present a new infinite-dimensional Lie group of intrinsic symmetries of anisotropic plasma equilibria equations, similar to those for the isotropic case.*

## *Contents.*

## 1. Isotropic and anisotropic MHD plasma equilibria

### 1.1. Isotropic plasma equilibrium equations.

The system of equations of ideal isotropic magnetohydrodynamic (MHD) equilibrium is

$$\text{div}(\rho \cdot \mathbf{V}) = 0,$$

$$\rho \cdot \mathbf{V} \times (\text{curl } \mathbf{V}) - \frac{1}{\mu} \cdot \mathbf{B} \times (\text{curl } \mathbf{B}) - \text{grad } P - \rho \cdot \text{grad } \frac{\mathbf{V}^2}{2} = 0, \quad \textbf{(1.1)}$$

$$\text{curl}(\mathbf{V} \times \mathbf{B}) = 0,$$

$$\text{div } \mathbf{B} = 0.$$

Here $\mathbf{V}$ is plasma velocity, $\mathbf{B}$ is magnetic field, $\rho$ - plasma density, $\mu$ - magnetic permeability coefficient, $P$ – plasma pressure. The coefficient $\mu$ can be removed by scaling $\mathbf{B}$, therefore from now on we assume $\mu=1$.

The most important reductions of this system are listed below. The system

$$(\text{curl } \mathbf{B}) \times \mathbf{B} = \text{grad } P, \quad \textbf{(1.2)}$$

$$\text{div } \mathbf{B} = 0$$

is a called Plasma Equilibrium equations; it is a reduction of (1.1) for the case of motionless plasma.

The system of force-free plasma equilibrium equations

$$\text{curl } \mathbf{B} = \alpha(\mathbf{x})\mathbf{B}, \quad \textbf{(1.3)}$$

$$\text{div } \mathbf{B} = 0,$$

is in turn a reduction of (1.2) when (curl $\mathbf{B}$) and $\mathbf{B}$ are collinear.

The simplest case of (1.3) is $\alpha(\mathbf{x})=0$, which gives rise to pure magnetic field equations

$$\text{curl } \mathbf{B} = 0, \quad \textbf{(1.4)}$$

$$\text{div } \mathbf{B} = 0.$$

All of the above reductions, as well as the original MHD equilibrium system, will be studied in this paper.

## 1.2. Anisotropic plasma equilibrium equations.

The system of equations of anisotropic magnetohydrodynamic (AMHD) equilibrium is

$$\text{div}(\rho \cdot \mathbf{V}) = 0,$$

$$\rho \cdot \mathbf{V} \times (\text{curl } \mathbf{V}) - \frac{1}{\mu} \cdot \mathbf{B} \times (\text{curl } \mathbf{B}) - \text{grad } \mathbf{P} - \rho \cdot \text{grad} \frac{V^2}{2} = 0, \quad (1.5)$$

$$\text{curl}(\mathbf{V} \times \mathbf{B}) = 0,$$

$$\text{div } \mathbf{B} = 0.$$

The coefficient $\mu$ can be removed by scaling $\mathbf{B}$, therefore we put $\mu=1$.

This system is analogous to (1.1) in everything except the fact that the pressure $\mathbf{P}$ is now a 3×3 tensor. In the case of small Larmor radius, pressure tensor has only two independent components:

$$\mathbf{P} = \mathbf{I} \cdot p_\perp + \frac{p_\parallel - p_\perp}{\mathbf{B}^2}(\mathbf{B}\mathbf{B}), \quad (1.6)$$

here $\mathbf{I}$ is unit tensor. Using vector calculus identities, one can represent gradient of tensor pressure as follows:

$$\text{grad } \mathbf{P} = \text{grad } p_\perp + \tau \cdot (\text{curl } \mathbf{B}) \times \mathbf{B} + \tau \cdot \text{grad} \frac{\mathbf{B}^2}{2} + \mathbf{B} \cdot (\mathbf{B} \cdot \text{grad } \tau), \quad (1.7)$$

$$\tau = \frac{p_\parallel - p_\perp}{\mathbf{B}^2}.$$

Hence the equation of conservation of momentum in (1.5) can be rewritten as

$$\rho \cdot \mathbf{V} \times (\text{curl } \mathbf{V}) - (1-\tau) \cdot \mathbf{B} \times (\text{curl } \mathbf{B}) =$$

$$\text{grad } p_\perp + \tau \cdot \text{grad} \frac{\mathbf{B}^2}{2} + \rho \cdot \text{grad} \frac{V^2}{2} + \mathbf{B} \cdot (\mathbf{B} \cdot \text{grad } \tau). \quad (1.8)$$

The motionless case of (1.8) is

$$(1-\tau) \cdot (\text{curl } \mathbf{B}) \times \mathbf{B} = \text{grad } p_\perp + \tau \cdot \text{grad} \frac{\mathbf{B}^2}{2} + \mathbf{B} \cdot (\mathbf{B} \cdot \text{grad } \tau). \quad (1.9)$$

## 2. Bäcklund transformation connecting solutions of MHD and AMHD equilibrium systems of equations.

Consider the system (1.1). Let $\Psi$ be a function constant both on the magnetic field lines and on the streamlines of plasma, and let the density $\rho = \rho(\Psi)$.

Suppose $\{\mathbf{B}, \mathbf{V}, \rho, P\}$ is a solution of (1.1). We introduce two vector fields

$$\mathbf{B}_1 = f(\Psi) \cdot \mathbf{B}; \quad \mathbf{V}_1 = g(\Psi) \cdot \mathbf{V}. \tag{2.1}$$

Let us now show that it is possible to make these new vector fields satisfy the AMHD equilibrium equations (1.5). Indeed,

$$\mathrm{div}\,\mathbf{B}_1 = f(\Psi) \cdot \mathrm{div}\,\mathbf{B} + \mathrm{grad}\,f(\Psi) \cdot \mathbf{B} = 0, \tag{2.2}$$

$$\mathrm{div}\,\mathbf{V}_1 = g(\Psi) \cdot \mathrm{div}\,\mathbf{V} + \mathrm{grad}\,g(\Psi) \cdot \mathbf{V} = 0, \tag{2.3}$$

$$\mathrm{curl}\,(\mathbf{V}_1 \times \mathbf{B}_1) = f \cdot g \cdot \mathrm{curl}\,(\mathbf{V} \times \mathbf{B}) + (f \cdot g)' \cdot \mathrm{grad}\,\Psi \times (\mathbf{V} \times \mathbf{B}) = 0, \tag{2.4}$$

We have to satisfy the requirement

$$\mathrm{div}\,(\rho_1 \mathbf{V}_1) = 0, \tag{2.5}$$

which can be done by choosing $\rho_1 = \rho_1(\Psi)$, and also the momentum conservation equation (1.8)

$$\rho_1 \cdot \mathbf{V}_1 \times (\mathrm{curl}\,\mathbf{V}_1) - (1 - \tau_1) \cdot \mathbf{B}_1 \times (\mathrm{curl}\,\mathbf{B}_1) =$$
$$\mathrm{grad}\,p_\perp + \tau_1 \cdot \mathrm{grad}\,\frac{\mathbf{B}_1^{\,2}}{2} + \rho_1 \cdot \mathrm{grad}\,\frac{\mathbf{V}_1^{\,2}}{2} + \mathbf{B}_1 \cdot (\mathbf{B}_1 \cdot \mathrm{grad}\,\tau_1). \tag{2.6}$$

Let us expand left- and right-hand sides of (2.6) separately. From (2.1),

$$\mathrm{curl}\,\mathbf{B}_1 = f(\Psi) \cdot \mathrm{curl}\,\mathbf{B} + \mathrm{grad}\,f(\Psi) \times \mathbf{B},$$

therefore

$$\mathbf{B}_1 \times \mathrm{curl}\,\mathbf{B}_1 = f^2 \cdot \mathbf{B} \times \mathrm{curl}\,\mathbf{B} + ff' \cdot \mathbf{B}^2 \cdot \mathrm{grad}\,\Psi, \tag{2.7}$$

$$\mathbf{V}_1 \times \mathrm{curl}\,\mathbf{V}_1 = g^2 \cdot \mathbf{V} \times \mathrm{curl}\,\mathbf{V} + gg' \cdot \mathbf{V}^2 \cdot \mathrm{grad}\,\Psi. \tag{2.8}$$

The left-hand side of (2.6) then takes the form

$$\rho_1 \cdot g^2 \cdot \mathbf{V} \times (\mathrm{curl}\,\mathbf{V}) - (1 - \tau_1) \cdot f^2 \cdot \mathbf{B} \times (\mathrm{curl}\,\mathbf{B}) + \left(\rho_1 \cdot gg' \cdot \mathbf{V}^2 - (1 - \tau_1) \cdot ff' \cdot \mathbf{B}^2\right) \cdot \mathrm{grad}\,\Psi,$$

and the right-hand side:

$$\mathrm{grad}\,p_\perp + \tau_1 \cdot \mathrm{grad}\,\frac{\mathbf{B}_1^{\,2}}{2} + \rho_1 \cdot \mathrm{grad}\,\frac{\mathbf{V}_1^{\,2}}{2} + \mathbf{B}_1 \cdot (\mathbf{B}_1 \cdot \mathrm{grad}\,\tau_1).$$

To make them agree, we make assumptions

$$\tau_1 = \tau_1(\Psi), \tag{2.9}$$

$$\frac{\rho_1 \cdot g^2}{(1 - \tau_1) \cdot f^2} = \rho, \tag{2.10}$$

Then, after using (1.1), the left-hand side of (2.6) takes the form

$$(1 - \tau_1) \cdot f^2 \cdot \left(\mathrm{grad}\,P + \rho \cdot \mathrm{grad}\,\frac{\mathbf{V}^2}{2}\right) + \left(\rho_1 \cdot gg' \cdot \mathbf{V}^2 - (1 - \tau_1) \cdot ff' \cdot \mathbf{B}^2\right) \cdot \mathrm{grad}\,\Psi,$$

or

$$(1-\tau_1) \cdot f^2 \cdot \text{grad}\left[P + \rho \cdot \frac{V^2}{2}\right] + \left(\left[\rho_1 \cdot gg' - (1-\tau_1) \cdot f^2 \cdot \frac{\rho_1'}{2}\right] \cdot V^2 - (1-\tau_1) \cdot ff' \cdot B^2\right) \cdot \text{grad } \Psi. \quad (2.11)$$

The right-hand side rewrites as

$$\text{grad}\left(p_\perp + \tau_1 \cdot \frac{f^2 \cdot B^2}{2} + \rho_1 \cdot \frac{g^2 \cdot V^2}{2}\right) - \frac{1}{2}\left(f^2 \cdot B^2 \cdot \tau_1' + g^2 \cdot V^2 \cdot \rho_1'\right) \cdot \text{grad } \Psi. \quad (2.12)$$

Velocity and magnetic field depend not only on the magnetic surface variable $\Psi$, therefore in order to make the (2.11) and (2.12) equal we must first demand

$$P + \rho \cdot \frac{V^2}{2} = M(\Psi), \quad p_\perp + \tau_1 \cdot f^2 \cdot \frac{B^2}{2} + \rho_1 \cdot g^2 \cdot \frac{V^2}{2} = N(\Psi), \quad N'(\Psi) = M'(\Psi) \cdot (1-\tau_1) \cdot f^2,$$

so

$$P = M(\Psi) - \rho \cdot \frac{V^2}{2}, \quad p_\perp = N(\Psi) - \tau_1 \cdot f^2 \cdot \frac{B^2}{2} - \rho_1 \cdot g^2 \cdot \frac{V^2}{2}.$$

Setting terms of (2.11) and (2.12) containing respectively velocity and magnetic field to be equal, we get

$$\frac{1}{2} f^2 \cdot \tau_1' = (1-\tau_1) \cdot ff', \quad -\frac{1}{2} g^2 \cdot \rho_1' = \rho_1 \cdot gg' - (1-\tau_1) \cdot f^2 \cdot \frac{\rho_1'}{2}. \quad (2.13)$$

The first condition from (2.13) results in a formula for $\tau_1$:

$$\tau_1 = 1 - \frac{C_0}{f^2}. \quad (2.14)$$

Using this, we calculate:

$$\rho_1 = C_0 \cdot \rho / g^2, \quad N'(\Psi) = M'(\Psi) \cdot C_0. \quad (2.15)$$

The second equation of (2.13) then becomes an identity, and thus we have established a transformation from a class of solutions of isotropic MHD equilibrium equations into a class of solutions of AMHD equilibrium equations.

This result can be formulated as a theorem.

**Theorem 2.1.**

Let $\{\mathbf{B}, \mathbf{V}, \rho, P\}$ be a solution to isotropic MHD equilibrium equations (1.1) with properties

$$\rho = \rho(\Psi), \quad P = M(\Psi) - \rho \cdot \frac{\mathbf{V}^2}{2},$$

where $\Psi$ is a function constant both on the magnetic field lines and on plasma streamlines, and $M$ is an arbitrary function of $\Psi$. Then the transformation

$$\mathbf{B}_1 = f(\Psi) \cdot \mathbf{B}, \quad \mathbf{V}_1 = g(\Psi) \cdot \mathbf{V},$$

$$\rho_1 = \frac{C_0 \cdot \rho}{g(\Psi)^2}, \quad \tau_1 = 1 - \frac{C_0}{f(\Psi)^2},$$

$$p_\perp = C_0 P + C_1 + \left(C_0 - f(\Psi)^2\right) \frac{\mathbf{B}^2}{2}$$

defines a solution to anisotropic MHD equilibrium equations (1.5)-(1.7). Here $f(\Psi)$, $g(\Psi)$ are arbitrary continuously differentiable functions, and $C_0$, $C_1$ are arbitrary constants.

## *3. Discussion and important cases.*

Theorem (2.1) can be used to build solutions of AMHD equilibrium and anisotropic plasma equilibrium equations, starting from isotropic MHD equilibrium, plasma equilibrium, and even vacuum magnetic field configurations. The theorems below describe exact procedures of such construction.

**Theorem 3.1.**

Let $\{\mathbf{B}, p\}$ be an arbitrary solution to isotropic plasma equilibrium equations (1.2). Then the transformations

$$\mathbf{B}_1 = f(p)\mathbf{B},$$

$$\tau_1 = 1 - \frac{C_0}{f(p)^2},$$

$$p_\perp = C_0 p + C_1 + \left(C_0 - f(p)^2\right)\frac{\mathbf{B}^2}{2}$$

define a solution $\{\mathbf{B}_1, p_\perp, \tau_1\}$ to anisotropic plasma equilibrium equations (1.9). Here $f(p)$ is an arbitrary continuously differentiable function of pressure $p$; $C_0$, $C_1$ are arbitrary constants.

**Theorem 3.2.**

Let $\mathbf{B}$ be an arbitrary solution to the equations for pure magnetic field in vacuum (1.4):

$$\operatorname{curl}\mathbf{B} = 0,$$

$$\operatorname{div}\mathbf{B} = 0,$$

and let $p$ be a function constant on magnetic field lines. Then the formulas

$$\mathbf{B}_1 = f(p)\mathbf{B},$$

$$\tau_1 = 1 - \frac{C_0}{f(p)^2}$$

$$p_\perp = C_1 + \left(C_0 - f(p)^2\right)\frac{\mathbf{B}^2}{2}$$

define a solution $\{\mathbf{B}_1, p_\perp, \tau_1\}$ to anisotropic plasma equilibrium equations (1.9). Here $f(p)$ is an arbitrary continuously differentiable function of $p$; $C_0$, $C_1$ are arbitrary constants.

Theorems 3.1, 3.2 can be applied to construct a wide variety of anisotropic plasma equilibrium solutions of different topologies. Indeed, with the help of Theorem 3.2, using any harmonic function $\phi$: $\Delta\phi = 0$ we can build an anisotropic plasma equilibrium.

## 4. Infinite-dimensional Lie group of transformations of anisotropic MHD equilibria.

The system of anisotropic magnetohydrodynamics (AMHD) equilibrium equations can be represented as (see (1.5), (1.7), (1.8))

$$\rho \cdot \mathbf{V} \times (\operatorname{curl} \mathbf{V}) - (1-\tau) \cdot \mathbf{B} \times (\operatorname{curl} \mathbf{B}) =$$
$$\operatorname{grad} p_\perp + \tau \cdot \operatorname{grad} \frac{\mathbf{B}^2}{2} + \rho \cdot \operatorname{grad} \frac{\mathbf{V}^2}{2} + \mathbf{B} \cdot (\mathbf{B} \cdot \operatorname{grad} \tau), \quad (4.1)$$

$$\operatorname{div}(\rho \cdot \mathbf{V}) = 0, \quad \operatorname{div} \mathbf{B} = 0, \quad (4.2)$$

$$\operatorname{curl}(\mathbf{V} \times \mathbf{B}) = 0, \quad (4.3)$$

$$\tau = \frac{p_\| - p_\perp}{\mathbf{B}^2}. \quad (4.4)$$

It is possible to show that these equations possess symmetries that are very similar to those for isotropic plasma equilibria found by O. I. Bogoyavlenskij in [1-2], and are indeed a natural generalization of them for the anisotropic case.

This new class of symmetries allows building rich families of AMHD equilibrium solutions from single known solutions, and the properties of the new solutions can be chosen appropriately in quite wide range, depending on an application.

**Theorem 4.1.**

Let $\{\mathbf{V}(\mathbf{r}), \mathbf{B}(\mathbf{r}), p_\perp(\mathbf{r}), \rho(\mathbf{r}), \tau(\mathbf{r})\}$ be a solution of (4.1)-(4.4), where the density $\rho(\mathbf{r})$ and the function $\tau(\mathbf{r})$ are constant on both magnetic field lines and streamlines. Then $\{\mathbf{V}_1(\mathbf{r}), \mathbf{B}_1(\mathbf{r}), p_{\perp 1}(\mathbf{r}), \rho_1(\mathbf{r}), \tau_1(\mathbf{r})\}$ is also a solution of the same system, where

$$\rho_1(\mathbf{r}) = m^2(\mathbf{r}) \cdot \rho(\mathbf{r}), \quad \tau_1(\mathbf{r}) = 1 - n^2(\mathbf{r}) \cdot (1 - \tau(\mathbf{r})), \quad (4.5)$$

$$\mathbf{V}_1(\mathbf{r}) = \frac{b(\mathbf{r}) \cdot \sqrt{1-\tau(\mathbf{r})}}{m(\mathbf{r}) \cdot \sqrt{\rho(\mathbf{r})}} \cdot \mathbf{B}(\mathbf{r}) + \frac{a(\mathbf{r})}{m(\mathbf{r})} \cdot \mathbf{V}(\mathbf{r}), \quad (4.6)$$

$$\mathbf{B}_1(\mathbf{r}) = \pm \left( \frac{a(\mathbf{r})}{n(\mathbf{r})} \cdot \mathbf{B}(\mathbf{r}) + \frac{b(\mathbf{r}) \cdot \sqrt{\rho(\mathbf{r})}}{n(\mathbf{r}) \cdot \sqrt{1-\tau(\mathbf{r})}} \cdot \mathbf{V}(\mathbf{r}) \right), \quad (4.7)$$

$$p_{\perp 1}(\mathbf{r}) = C \cdot p_\perp(\mathbf{r}) + (C\mathbf{B}(\mathbf{r})^2 - \mathbf{B}_1(\mathbf{r})^2)/2. \quad (4.8)$$

Here $a(\mathbf{r})$, $b(\mathbf{r})$, $m(\mathbf{r})$, $n(\mathbf{r})$ are arbitrary differentiable functions constant on both magnetic field lines and streamlines (i.e. on magnetic surfaces $\Psi$ = const, if they exist); $a(\mathbf{r})$ and $b(\mathbf{r})$ satisfying $a^2(\mathbf{r}) - b^2(\mathbf{r}) = C = \operatorname{const}$.

**Proof of the theorem.**

By the assumption of the theorem, functions $a(\mathbf{r})$, $b(\mathbf{r})$, $m(\mathbf{r})$, $n(\mathbf{r})$, $\rho(\mathbf{r})$, $\tau(\mathbf{r})$ are constant on streamlines and magnetic field lines. Using this fact and the identities (4.5)-(4.8), we can verify that equations (4.2) hold:

$$\text{div}(\rho_1 \cdot \mathbf{V}_1) = \text{grad}\left(m(\mathbf{r}) \cdot \sqrt{\rho(\mathbf{r})} \cdot b(\mathbf{r}) \cdot \sqrt{1-\tau(\mathbf{r})}\right) \cdot \mathbf{B} + \left(m(\mathbf{r}) \cdot \sqrt{\rho(\mathbf{r})} \cdot b(\mathbf{r}) \cdot \sqrt{1-\tau(\mathbf{r})}\right) \cdot \text{div}\,\mathbf{B}$$

$$+ \text{grad}\left(m(\mathbf{r}) \cdot \rho(\mathbf{r}) \cdot a(\mathbf{r})\right) \cdot \mathbf{V} + \left(m(\mathbf{r}) \cdot \rho(\mathbf{r}) \cdot a(\mathbf{r})\right) \cdot \text{div}\,\mathbf{V} = 0,$$

$$\text{div}\,\mathbf{B}_1 = \pm\left(\text{grad}\frac{a(\mathbf{r})}{n(\mathbf{r})} \cdot \mathbf{B} + \frac{a(\mathbf{r})}{n(\mathbf{r})} \cdot \text{div}\,\mathbf{B} + \text{grad}\frac{b(\mathbf{r}) \cdot \sqrt{\rho(\mathbf{r})}}{n(\mathbf{r}) \cdot \sqrt{1-\tau(\mathbf{r})}} \cdot \mathbf{V} + \frac{b(\mathbf{r}) \cdot \sqrt{\rho(\mathbf{r})}}{n(\mathbf{r}) \cdot \sqrt{1-\tau(\mathbf{r})}} \cdot \text{div}\,\mathbf{V}\right) = 0.$$

The equation (4.3) is verified in the same manner:

$$\text{curl}\,(\mathbf{V}_1 \times \mathbf{B}_1) = \pm\text{curl}\left[(\mathbf{V} \times \mathbf{B}) \cdot \left(\frac{a(\mathbf{r})}{m(\mathbf{r})} \cdot \frac{a(\mathbf{r})}{n(\mathbf{r})} - \frac{b(\mathbf{r}) \cdot \sqrt{1-\tau(\mathbf{r})}}{m(\mathbf{r}) \cdot \sqrt{\rho(\mathbf{r})}} \cdot \frac{b(\mathbf{r}) \cdot \sqrt{\rho(\mathbf{r})}}{n(\mathbf{r}) \cdot \sqrt{1-\tau(\mathbf{r})}}\right)\right]$$

$$= \pm\left(\text{curl}(\mathbf{V} \times \mathbf{B})\right) \cdot \left(\frac{a^2(\mathbf{r}) - b^2(\mathbf{r})}{m(\mathbf{r}) \cdot n(\mathbf{r})}\right) \pm (\mathbf{V} \times \mathbf{B}) \times \text{grad}\left(\frac{a^2(\mathbf{r}) - b^2(\mathbf{r})}{m(\mathbf{r}) \cdot n(\mathbf{r})}\right) = 0.$$

This is true because the vectors $(\mathbf{V} \times \mathbf{B})$ and $\text{grad}\left(\dfrac{a^2(\mathbf{r}) - b^2(\mathbf{r})}{m(\mathbf{r}) \cdot n(\mathbf{r})}\right)$ are both orthogonal to magnetic field lines and streamlines (which do not coincide), and therefore parallel. (Note: in the case $\mathbf{V} \parallel \mathbf{B}$, when the previous statement is incorrect, $(\mathbf{V} \times \mathbf{B}) = 0$ identically).

We also have to verify that equation (4.1) holds for the transformed plasma parameters. It. can be rewritten as

$$\rho_1 \cdot \mathbf{V}_1 \times (\text{curl}\,\mathbf{V}_1) - (1-\tau_1) \cdot \mathbf{B}_1 \times (\text{curl}\,\mathbf{B}_1)$$

$$= \rho_1 \cdot \text{grad}\frac{\mathbf{V}_1^2}{2} - (1-\tau_1) \cdot \text{grad}\frac{\mathbf{B}_1^2}{2} + \text{grad}\left(p_{\perp 1} + \frac{\mathbf{B}_1^2}{2}\right). \quad (4.9)$$

We denote $\mathbf{A}_1 = \sqrt{\rho_1} \cdot \mathbf{V}_1$, $\mathbf{M}_1 = \sqrt{1-\tau_1} \cdot \mathbf{B}_1$, where

$$\mathbf{A}_1 = b(\mathbf{r}) \cdot \sqrt{1-\tau(\mathbf{r})} \cdot \mathbf{B} + a(\mathbf{r}) \cdot \sqrt{\rho(\mathbf{r})} \cdot \mathbf{V}, \quad (4.10)$$

$$\mathbf{M}_1 = \pm\left(a(\mathbf{r}) \cdot \sqrt{1-\tau(\mathbf{r})} \cdot \mathbf{B} + b(\mathbf{r}) \cdot \sqrt{\rho(\mathbf{r})} \cdot \mathbf{V}\right). \quad (4.11)$$

It is known that for any function $f(\mathbf{r})$ and a vector field $\mathbf{a}$ the following relation holds:

$$(f\mathbf{a}) \times \left[\text{curl}\,(f\mathbf{a})\right] = f^2 \cdot \mathbf{a} \times (\text{curl}\,\mathbf{a}) + \frac{\mathbf{a}^2}{2} \cdot \text{grad}\,f^2 - f \cdot (\mathbf{a} \cdot \text{grad}\,f) \cdot \mathbf{a}. \quad (4.12)$$

If we write the equation (4.9) in terms of $\mathbf{A}_1$ and $\mathbf{M}_1$ and use (4.12), it takes the form

$$\mathbf{A}_1 \times (\text{curl}\,\mathbf{A}_1) - \mathbf{M}_1 \times (\text{curl}\,\mathbf{M}_1) = \text{grad}\left(p_{\perp 1} + \frac{\mathbf{A}_1^2}{2} - \frac{\mathbf{M}_1^2}{2} + \frac{\mathbf{B}_1^2}{2}\right). \quad (4.13)$$

Substituting here the formulae (4.5)-(4.8), (4.10)-(4.11), we get

$$(a^2 - b^2) \cdot \rho \cdot \mathbf{V} \times (\text{curl } \mathbf{V}) - (a^2 - b^2) \cdot (1 - \tau) \cdot \mathbf{B} \times (\text{curl } \mathbf{B})$$

$$+ \frac{\mathbf{V}^2}{2} \text{grad} \left[ (a^2 - b^2) \cdot \rho \right] - \frac{\mathbf{B}^2}{2} \text{grad} \left[ (a^2 - b^2) \cdot (1 - \tau) \right] \tag{4.14}$$

$$= \text{grad} \left( C \cdot p_\perp(\mathbf{r}) + C \frac{\mathbf{B}^2}{2} + \frac{\mathbf{A}_1^2}{2} - \frac{\mathbf{M}_1^2}{2} \right).$$

Using the identity $a^2(\mathbf{r}) - b^2(\mathbf{r}) = C$, the fact $\mathbf{A}_1^2 - \mathbf{M}_1^2 = C \cdot \left( \rho \cdot \mathbf{V}^2 - (1 - \tau) \cdot \mathbf{B}^2 \right)$ and the formula (4.1) which is satisfied by $\{\mathbf{V}(\mathbf{r}), \mathbf{B}(\mathbf{r}), p_\perp(\mathbf{r}), \rho(\mathbf{r}), \tau(\mathbf{r})\}$, we transform (4.14) to the equivalent form

$$C \cdot \left[ \text{grad } p_\perp + \tau \cdot \text{grad } \frac{\mathbf{B}^2}{2} + \rho \cdot \text{grad } \frac{\mathbf{V}^2}{2} \right] + C \cdot \frac{\mathbf{V}^2}{2} \text{grad } \rho - C \cdot \frac{\mathbf{B}^2}{2} \text{grad } (1 - \tau)$$

$$= \text{grad} \left( C \cdot p_\perp(\mathbf{r}) + C \frac{\mathbf{B}^2}{2} + \frac{C \cdot \left( \rho \cdot \mathbf{V}^2 - (1 - \tau) \cdot \mathbf{B}^2 \right)}{2} \right), \tag{4.14}$$

which is an identity. Theorem is proven.

## *References.*

1. O. I. Bogoyavlenskij. Phys. Rev. E 62 (6), 8616-8627 (2000).
2. O. I. Bogoyavlenskij. Phys. Lett. A 291, 256 (2001).